# An Optimized Design of Reversible Sequential Digital Circuits

Pradeep Singla [1] , Aakash Gupta [2] , Ashutosh Bhardwaj[3], Pulkit Basia[4]

*[1,3,4] Department of Electronics and Communication Engineering,* [2]*Deptt. Of Computer Science*
*Sonipat Institute of Engineering & Mgmt.*

*[1]pradeepsingla7@gmail.com*
*[2]aakashgarg1987@gmail.com*
*[3]ashu22bhardwaj@gmail.com*
*[4]pulkitbasia1292@gmail.com*

Abstract-   **In the today's era, reversible logics are the promising technology for the designing of low power digital logic system having major application in the field of nanotechnology, quantum computation, DNA and other low power digital circuits. Reversible logics provide zero power dissipation (Ideally) in the digital operations. There are numbers of circuit designed by the reversible logics and sequential circuits have their own importance in the digital systems. In this paper authors provides a optimized approach and optimized design for the sequential circuit ( counter as an example) by using the MUX gate ( a reversible gate) which provides the better results against the previous designs discussed in the literature. The proposed design has lower quantum cost, garbage output, constant input and total number of logical calculations performing by the design.**

## I. Introduction

Energy dissipation is one of the important considerations in hardware design. Reversible logic was primarily related to energy when in 1961 Landauer states that losing of a bit in digital circuits causes an smallest amount of heat dissipation and the theoretical limit of heat dissipation for losing of one bit computation would be KTln2 [1] where K is Boltzmann's constant and T is the temperature . In 1973 Bennett proved that KTln2 energy would not dissipate from a system as long as system allows the reproduction of the inputs from the observed outputs [2].Zero energy dissipation would be possible only if the network consist of reversible gates. In a reversible circuit input vector can be uniquely recovered from the output vector and here is one to one correspondence between inputs and outputs. Energy dissipation can be reduced if computation becomes information-lossless [3].The idea of reversible computing comes from the Thermodynamics, which taught us the benefits of reversible process over irreversible process. These circuits can generate unique output vector from each input vector and vice versa [4].

Reversible logic synthesis of reversible sequential logic differs from combinational logic in that the output of the logic device is dependent not only on the present inputs to the device, but also on past inputs. In this paper we are synthesising on sequential logics by using reversible gates and the paper provides the optimized design of 3 bit synchronous and asynchronous counter in terms of garbage outputs, number of total calculations, quantum cost[5]. The reversible MUX gate and Feynman gate is used for designing the J-K flip-flop which is an essential component of 3 bit counter. The numeral results are shown later.

This paper presents following prospective: Section ll describe the background of the reversible logic and the conditions for the reversibility. Different reversible gate structure is also discussed in the same section. In Section III, we have discussed basic necessities of counter and its following types. In section IV, we have proposed design of three bit reversible synchronous and asynchronous counter. In the Section V, we have described numerical results of proposed reversible synchronous and asynchronous counter.

## II. REVERSIBLE LOGIC

The logical reversibility means there should be same number of input and outputs i.e. The number of input lines and output lines must be same and there must be one to one mapping between input and output[3].The gate must run forward and backward i.e., the input can also be retrieved from the output, when device obeys these two conditions then the second law of thermodynamics guarantees that it dissipate no heat[4].In the synthesis of reversible circuits nor direct fan out is allowed neither feedback is permitted. A reversible circuit should be designed using minimum number of reversible logic gates. From the point of view of reversible circuit design there are parameter for determining the complexity and performance of circuit, such as garbage output and quantum cost [6]. Some important terms used in reversible logics are as follows;

- Garbage output[3] -Garbage output refers to the number of unused outputs present in a reversible logic circuit.
- Quantum cost[5]-Quantum cost refers to the cost of circuits in terms of the cost of primitive gates.
- Total Logical calculation[3] - The Total logical calculation is the count of the XOR, AND, NOT logic in the output circuit.
- Constant inputs[3]–Constant inputs is the number of inputs that are to be maintained constant at either 0 or 1 in order to synthesize the given logical function

### A. Basic Reversible Logic Gates–

#### 1) Feynman Gate:

It is a 2X2 Feynman gate having two inputs A and B and two outputs P and Q. The outputs are defined by P=A, Q=A xor B. Quantum cost of a Feynman gate is 1. Figure 1 shows a 2X2 Feynman gate[3][4][5][7] and Table I shows truth table of Feynman gate.

Table 1

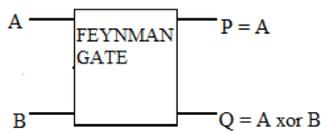

Figure1. Feynman gate

| A | B | C | D |
|---|---|---|---|
| 0 | 0 | 0 | 0 |
| 0 | 1 | 0 | 1 |
| 1 | 0 | 1 | 1 |
| 1 | 1 | 1 | 0 |

#### 2) Fredkin Gate:

It is a 3X3 Fredkin gate having three inputs A, B and C and three outputs P, Q, and R. The output is defined by P=A, Q= A'B xor AC and R= A'C xor AB. Quantum cost of a Fredkin gate is 5. Figure 2 shows a 3X3 Fredkin gate [3][4][5][8]. Table II shows truth table of Fredkin Gate.

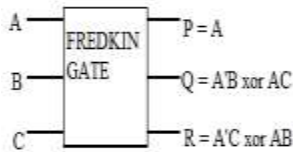

Figure 2. Fredkin gate

| A | B | C | P | Q | R |
|---|---|---|---|---|---|
| 0 | 0 | 0 | 0 | 0 | 0 |
| 0 | 0 | 1 | 0 | 0 | 1 |
| 0 | 1 | 0 | 0 | 1 | 0 |
| 0 | 1 | 1 | 0 | 1 | 1 |
| 1 | 0 | 0 | 1 | 0 | 0 |
| 1 | 0 | 1 | 1 | 1 | 0 |
| 1 | 1 | 0 | 1 | 0 | 1 |
| 1 | 1 | 1 | 1 | 1 | 1 |

*3) MUX Gate :*

It is a 3X3 MUX gate having three inputs A,B and C and three outputs P,Q, and R. The output is defined by P=A, Q= A xor B xor C and R= A'C xor AB. Quantum cost of a MUX gate is 4. Figure 3 shows a 3X3 MUX gate [3]. Table III shows truth table of MUX Gate.

TABLE III

| A | B | C | P | Q | R |
|---|---|---|---|---|---|
| 0 | 0 | 0 | 0 | 0 | 0 |
| 0 | 0 | 1 | 0 | 1 | 1 |
| 0 | 1 | 0 | 0 | 1 | 0 |
| 0 | 1 | 1 | 0 | 0 | 1 |
| 1 | 0 | 0 | 1 | 1 | 0 |
| 1 | 0 | 1 | 1 | 0 | 0 |
| 1 | 1 | 0 | 1 | 0 | 1 |
| 1 | 1 | 1 | 1 | 1 | 1 |

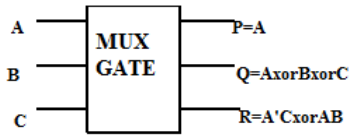

Figure 3. Mux gate

### III. COUNTER THEORY

Counter is a sequential machine constructed with the help of flip-flop. The counting of counter may be ascending, descending or in any manner decided by the designer[9][10].They are generally used in applications like digital alarm clock, managing production quantities, filling fixed quantities, time interval measurement etc..These devices generate binary numbers in a specified counting sequence when triggered by an incoming clock. After each trigger, the counter advances to the next number in the sequence. Counters may be used to generate waveforms of particular patterns and frequencies [10]. There are two kinds of counters in literature as stated below

*A. Asynchronous counter-*

In asynchronous counter the output of one flip-flop is feeded to the clock input of another flip-flop[11]. The control clock is input to the first stage. Figure 4 shows three bit asynchronous counter.

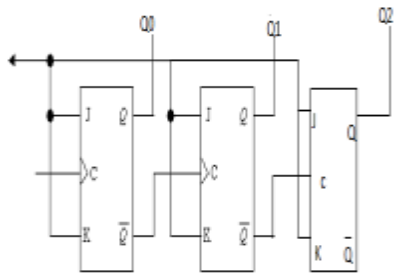

Fig 4.Three bit asynchronous counter

*B. Synchronous counter-*

Synchronous counter is a counter in which same clock pulse is applied to all flip-flops. Synchronous counters can be designed to count up and down in a specific order. They may be used to produce count sequences of non-consecutive numbers[12]. The count sequence produced by synchronous counters is not dependent on the triggering characteristics of the flip-flops[13]. Figure 5 shows a three-bit synchronous counter where;

J2 = Q1Q0, K2= Q1Q0, J1= Q0, K1= Q0, J0=1, K0=1

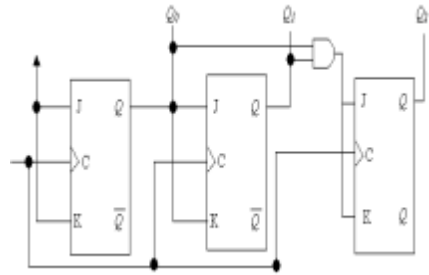

Figure 5.Three bit synchronous counter

## IV  PROPOSED REVERSIBLE THREE BIT SYNCHRONOUS AND ASYNCHRONOUS COUNTER

The reversible J-K flip flop(as shown in fig.9) is used to implement more complex reversible sequential circuits. Figure 6 shows three bit reversible synchronous counter implemented using J-K flip flop and Figure 7 shows three bit reversible asynchronous counter implemented using J-K flip flop with common clock input.

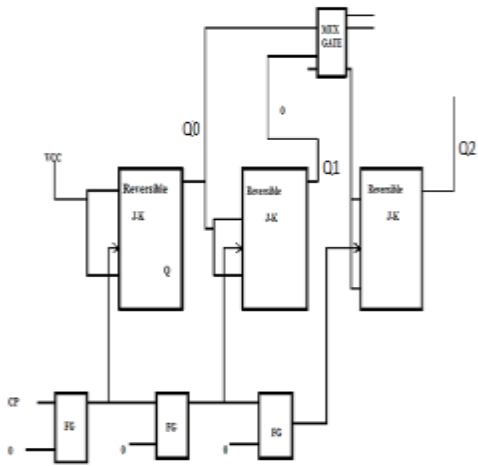

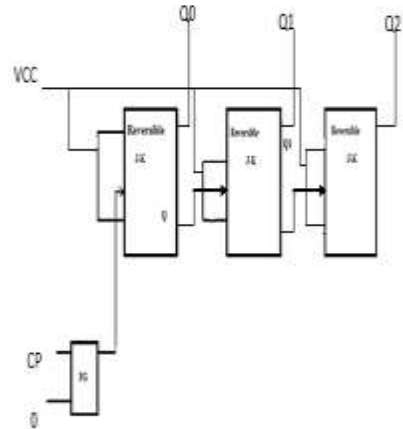

Fig 6. Three bit reversible synchronous counter

Fig 7.Three bit reversible Asynchronous counter

In three bit synchronous counter the clock is being passed through Feynman gate when Feynman gate is working in copy mode and another inputs are being fed through Vcc which is positive i.e.; now according to J-K flip flop its working would be in toggle condition, so if it is assumed that flip flop is in reset condition then due to toggle it will change its state to set mode i.e. 1.The cycle will run this way from reset to set & set to reset and so on.

# V. NUMERAL RESULT

*A . Quantum cost-*

*1). Quantum cost of reversible J-K flip-flop:*

In reversible J-K flip-flop we have used 4 Mux gate ,2 New gate and 4 Feynman gate .Now let m is the Quantum cost of Mux gate , n is the Quantum cost of the new gate and F is the Quantum cost of the Feynman gate . so, the Quantum cost(A) of the Reversible J-K flip-flop is

Quantum cost(A) = 4m+2n+4F

*2) Quantum cost of 3 bit reversible synchronous counter:*

In three bit reversible synchronous counter we have used 3 reversible J-K flip flop , 1 mux gate and 3 Feynman gate .Now let A be Quantum cost of reversible J-K flip-flop , m be the Quantum cost of the mux gate and F is the Quantum cost of Feynman gate . So, the Quantum cost of 3 bit reversible synchronous counter is

Quantum cost (C)= 3A+1m+3F = 3(4m+2n+4F)+1m+3F =13m+6n+15F

*3) Quantum cost of 3 bit reversible asynchronous counter:*

In three bit reversible asynchronous counter we have used 3 reversible J-K flip flops and 1 Feynman gate .Now let A be Quantum cost of reversible J-K flip-flop and F is the Quantum cost of Feynman gate. So, the Quantum cost of 3 bit reversible asynchronous counter is

Quantum cost (ac) = 3A+1F =3(4m+2n+4F) + 1F = 12m+6n+13f

*B . Total Logical calculation (T) :*

Assuming
$\alpha$ = A two input XOR gate calculation
$\beta$ = A two input AND gate calculation
$\delta$ = A NOT gate calculation
T = Total logical calculation
The Total logical calculation is the count of the XOR, AND, NOT logic in the output circuit. For example MUX gate has three XOR gate and two AND gate and one NOT gate in the output expression. Therefore ($\square$) = $3\alpha+2\beta+\delta$.and New gate has two XOR gate and two AND gate and three NOT gate in the output expression .therefore T (N)=$2\alpha+2\beta+3\delta$

*1) Total Logical calculation (T) of reversible J-K flip-flop:*

In reversible J-K flip-flop we have used 4 Mux gate, 2Newgates and 4 Feynman gate
So Total logical calculation of reversible J-K flip-flop is
T = 4× $(3\alpha+2\beta+\delta)$( for MUX gate)+ 4×1$\alpha$(for Feynman gate) +2(2 $\alpha+2\beta+3\delta$) (for New gate)

T (J-k) = 20 $\alpha$ +12 $\beta$ +10 $\delta$

*2) Total Logical calculation (T) of 3 bit reversible synchronous counter:*

In three bit reversible synchronous counter we have used 3 reversible J-K flip flop , 1 mux gate and 3 Feynman gate .So, Total Logical calculation (T) of 3 bit reversible synchronous counter
T =3 x (20 $\alpha$ +12 $\beta$ +10 $\delta$) +1× $(3\alpha+2\beta+\delta)$ (for MUX gate) +     3×1 $\alpha$ (for Feynman gate)
T=66 $\alpha$ +38 $\beta$ +31 $\delta$

*3) Total Logical calculation (T) of 3 bit reversible asynchronous counter:*

In three bit reversible asynchronous counter we have used 3 reversible J-K flip flops and 1 Feynman gate .So, Total Logical calculation (T) of 3 bit reversible synchronous counter.

$$T = 3x (20\,\alpha + 12\,\beta + 10\,\delta) + \times 1\alpha \ \text{(for Feynman gate)}$$
$$T = 61\,\alpha + 36\,\beta + 30\,\delta$$

TABLE  IV COMPARISON RESULTS

| Type | Design | Quantum cost | logical calculation |
|------|--------|--------------|---------------------|
| Proposed circuit | (Synchronous counter) | 109 | $66\alpha + 38\beta + 31\delta$ |
| | Asynchronous counter | 103 | $61\alpha + 36\beta + 30\delta$ |
| Existing one | (Synchronous counter) | 122 | $53\alpha + 64\beta + 45\delta$ |
| | Asynchronous counter | 115 | $49\alpha + 60\beta + 42\delta$ |

Where :
$\alpha$ = A two input XOR gate calculation
$\beta$ = A two input AND gate calculation
$\delta$ = A NOT gate calculation

## VI. CONCLUSION

The counter are mostly used for applications like filling fixed quantities, managing production quantities ,digital alarm clock, computer memory pointer, time interval measurement.
In this paper we have proposed a new improved design of three bit synchronous and asynchronous counter with the help of MUX gate and reversible JK flip-flops. The mux gate has lower quantum cost as compared to the Fredkin gate defined in literature and mux gate can all the operation as do as Fredkin. The numeral results are also shown in the paper with Table IV which shows the optimized results of proposed design against the previous designs of sequential circuits and will provides a new arena to design digital logical systems.

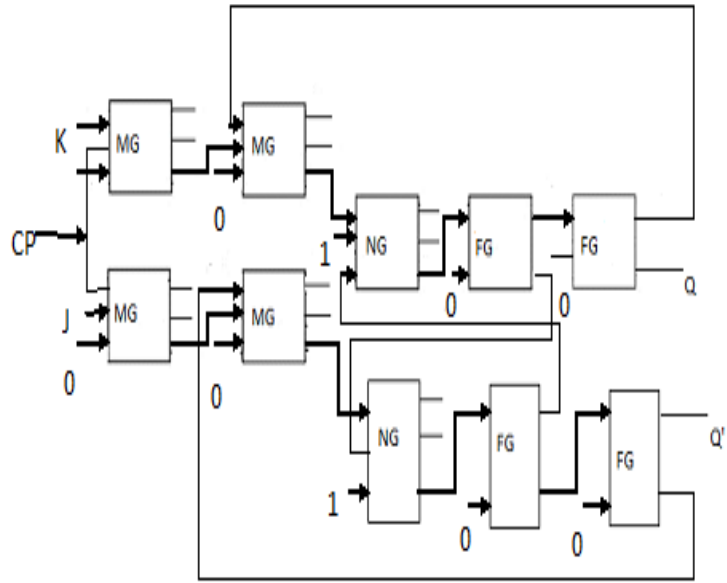

Fig 9 . Reversible JK Flip Flop